\PassOptionsToPackage{unicode}{hyperref}
\PassOptionsToPackage{hyphens}{url}
\documentclass[
  a4paper,
]{article}
\usepackage{amsmath,amssymb}
\usepackage{lmodern}
\usepackage{ifxetex,ifluatex}
\ifnum 0\ifxetex 1\fi\ifluatex 1\fi=0 
  \usepackage[T1]{fontenc}
  \usepackage[utf8]{inputenc}
  \usepackage{textcomp} 
\else 
  \usepackage{unicode-math}
  \defaultfontfeatures{Scale=MatchLowercase}
  \defaultfontfeatures[\rmfamily]{Ligatures=TeX,Scale=1}
\fi
\IfFileExists{upquote.sty}{\usepackage{upquote}}{}
\IfFileExists{microtype.sty}{
  \usepackage[]{microtype}
  \UseMicrotypeSet[protrusion]{basicmath} 
}{}
\makeatletter
\@ifundefined{KOMAClassName}{
  \IfFileExists{parskip.sty}{%
    \usepackage{parskip}
  }{
    \setlength{\parindent}{0pt}
    \setlength{\parskip}{6pt plus 2pt minus 1pt}}
}{
  \KOMAoptions{parskip=half}}
\makeatother
\usepackage{xcolor}
\IfFileExists{xurl.sty}{\usepackage{xurl}}{} 
\IfFileExists{bookmark.sty}{\usepackage{bookmark}}{\usepackage{hyperref}}
\hypersetup{
  pdftitle={Users' ability to perceive misinformation: An information quality assessment approach},
  pdfauthor={Aljaž Zrnec, Marko Poženel and Dejan Lavbič},
  hidelinks,
  pdfcreator={LaTeX via pandoc}}
\usepackage[margin=1in]{geometry}
\usepackage{longtable,booktabs,array}
\usepackage{calc} 
\usepackage{etoolbox}
\makeatletter
\patchcmd\longtable{\par}{\if@noskipsec\mbox{}\fi\par}{}{}
\makeatother
\IfFileExists{footnotehyper.sty}{\usepackage{footnotehyper}}{\usepackage{footnote}}
\makesavenoteenv{longtable}
\usepackage{graphicx}
\makeatletter
\def\maxwidth{\ifdim\Gin@nat@width>\linewidth\linewidth\else\Gin@nat@width\fi}
\def\maxheight{\ifdim\Gin@nat@height>\textheight\textheight\else\Gin@nat@height\fi}
\makeatother
\setkeys{Gin}{width=\maxwidth,height=\maxheight,keepaspectratio}
\makeatletter
\def\fps@figure{htbp}
\makeatother
\providecommand{\tightlist}{%
  \setlength{\itemsep}{0pt}\setlength{\parskip}{0pt}}
\usepackage{comment}
\usepackage[none]{hyphenat}
\usepackage[ruled,linesnumbered]{algorithm2e}

\SetCommentSty{mycommfont}

\usepackage{booktabs}
\usepackage{longtable}
\usepackage{array}
\usepackage{multirow}
\usepackage{wrapfig}
\usepackage{float}
\usepackage{colortbl}
\usepackage{pdflscape}
\usepackage{tabu}
\usepackage{threeparttable}
\usepackage{threeparttablex}
\usepackage[normalem]{ulem}
\usepackage{makecell}
\usepackage{xcolor}
\ifluatex
  \usepackage{selnolig}  
\fi
\usepackage[]{natbib}
\title{Users' ability to perceive misinformation: An information quality assessment approach}
\author{Aljaž Zrnec, Marko Poženel and Dejan Lavbič}
\date{}

\begin{document}
\maketitle

\begin{quote}
Aljaž Zrnec, Marko Poženel and \textbf{Dejan Lavbič}. 2022. \href{https://doi.org/10.1016/j.ipm.2021.102739}{\textbf{Users' ability to perceive misinformation: An information quality assessment approach}}, \href{https://www.sciencedirect.com/journal/information-processing-and-management}{Information Processing \& Management \textbf{(IPM)}}, 59(1).
\end{quote}

\hypertarget{abstract}{%
\section*{Abstract}\label{abstract}}
\addcontentsline{toc}{section}{Abstract}

Digital information exchange enables quick creation and sharing of information and thus changes existing habits. Social media is becoming the main source of news for end-users replacing traditional media. This also enables the proliferation of fake news, which misinforms readers and is used to serve the interests of the creators. As a result, automated fake news detection systems are attracting attention. However, automatic fake news detection presents a major challenge; content evaluation is increasingly becoming the responsibility of the end-user. Thus, in the present study we used information quality (IQ) as an instrument to investigate how users can detect fake news. Specifically, we examined how users perceive fake news in the form of shorter paragraphs on individual IQ dimensions. We also investigated which user characteristics might affect fake news detection. We performed an empirical study with 1123 users, who evaluated randomly generated stories with statements of various level of correctness by individual IQ dimensions. The results reveal that IQ can be used as a tool for fake news detection. Our findings show that (1) domain knowledge has a positive impact on fake news detection; (2) education in combination with domain knowledge improves fake news detection; and (3) personality trait conscientiousness contributes significantly to fake news detection in all dimensions.

\hypertarget{keywords}{%
\section*{Keywords}\label{keywords}}
\addcontentsline{toc}{section}{Keywords}

Fake news, Misinformation, Information quality, Big Five personality traits, Fake news detection, Case study

\hypertarget{introduction}{%
\section{Introduction}\label{introduction}}

The rapid integration of digitalisation in our lives increases the volume of data in the global data sphere and changes users' behaviour. The availability of various multimedia devices, constant connection to the internet, and overflowing information sources affect the way people access existing media \citep{chang_role_2020}. Traditional media are facing challenges accessing users in this new reality. Users are moving away from traditional to social media when seeking information. The reasons for this change are the possibilities that online information access offers, including timeliness, ease of access and an endless supply of knowledge and information. Undoubtedly, social media has significantly contributed to this transformation \citep{bondielli_survey_2019}. People can follow official social media channels and post on their networks, while journalists may use social media to gain public opinions on certain current topics and even discover potential new stories \citep{bondielli_survey_2019}. \citet{allcott_social_2017} state that \(62 \%\) of US adults obtain news on social media. Also the case study for Britain shows the growing importance of social media and the increasing share of social media news consumption \citep{newman_social_2013}.

The way news spreads on social media is different from its spread via former media technologies. Content may reach users without prior filtering, fact-checking, or editorial judgement, while news can be created online faster and cheaper. Unfortunately, that makes social media a possible abundant source of unverified and/or false information. The attractiveness of spreading misinformation lies in the fact that Websites and individuals may profit from spreading misinformation. For example, teenagers from Macedonia became wealthy by penny-per-click advertising during U.S. presidential elections \citep{zhou_survey_2020}. False news is seen as a great threat to public trust, the economy, and journalism. The problem of misinformation detection is a challenging task since misinformation appear in different forms and is made seemingly real.

In the literature, many approaches for misinformation detection have been proposed \citep{bondielli_survey_2019, zhou_survey_2020}. Due to the rapid diffusion of misinformation in social media and limited scalability of manual detection methods, automatic fact-checking methods gained the attention of the researchers \citep{shu_fake_2017, gravanis_behind_2019}. However, the automatic detection of fake news based on content is non-trivial. Solutions are usually specialized, and more importantly, do not involve comprehensive human knowledge of the domain. Besides regular news, a large body of fake news can also be created and disseminated for a variety of purposes, such as financial or political gain \citep{shu_fake_2017}, and can have significant effects on society and public opinion \citep{zhang_overview_2020}.

The terms false information and fake news may get confused with various other terms associated with inaccurate information. However, for online content, the term ``fake news'' has been de-facto used to denote false information in mainstream media \citep{bondielli_survey_2019}. Existing studies often use the term fake news with related concepts. \citet{wu_misinformation_2019} defined misinformation as an umbrella term to include all false or inaccurate information that is spread in social media. They organize different type of information into these key categories: misinformation \citep{kucharski_study_2016, wardle_fake_2017}, disinformation \citep{kshetri_economics_2017, wardle_fake_2017}, fake news \citep{shu_fake_2017}, rumour \citep{zubiaga_detection_2018}, urban legend, spam and troll. These terms all share a characteristic that the inaccurate messages can cause distress and various kinds of destructive effect through social media, especially when timely intervention is absent \citep{wu_misinformation_2019}. They also introduce the concept of unverified information, which can be false or true and accurate. \citet{zhou_survey_2020} also presented similar categorization of concepts that frequently co-occur or have overlapped with fake news: deceptive news \citep{allcott_social_2017, lazer_science_2018, shu_fake_2017}, false news \citep{vosoughi_spread_2018}, satire news \citep{conroy_automatic_2015, c_tandoc_jr_defining_2018, wardle_fake_2017}, disinformation, misinformation, cherry-picking \citep{asudeh_detecting_2020}, clickbait \citep{chen_misleading_2015}, and rumour \citep{zubiaga_detection_2018}.

Misinformation and disinformation both refer to fake or inaccurate information. However, disinformation is deliberately created to mislead the user while misinformation is not \citep{wu_misinformation_2019}. Despite the popularity of fake news, there has been no universal definition for the term. In literature, \citet{wu_misinformation_2019} define fake news as intentionally spread misinformation that is in the format of news. \citet{zhou_survey_2020} provide a broad definition of fake news as false news, and a narrow definition as intentionally false news published by a news outlet. In this paper, we will adopt the broad definition of fake news presented in \citet{zhou_survey_2020}. Our approach focuses on the ability to detect fake news for a regular user, based on information quality that the source contains.

Existing studies that explore fake news on social media mainly focus on its automatic detection. The rising quantity of fake news on social media and its rapid dissemination, make automatic detection approaches attractive \citep{bindu_discovering_2018}. Relatively few studies have investigated fake news detection from the user's (news consumer's) perspective. In social media, fake news can take several forms and shapes and is becoming more and more similar to proper news, making it more difficult to efficiently detect, both manually and automatically \citep{bondielli_survey_2019}. With the growing exposure of end-users to unverified information, users become indirectly responsible for content assessment. From this observation, we argue that it is worthwhile to investigate how successful users are in detecting fake news.

Our approach relies on an individual acting as a self fact-checker. Compared to expert-based fact-checking, an individual has less knowledge of a specific topic or domain and leads to less accurate results. It also poorly scales with volume due to the fatigue as a result of effort someone has to invest to critically evaluate contents. However, cautious users can employ the services of expert-based or crowd-sourced fact-checking websites (like Politifact, or Fiskitt respectively) and verify ground truth for the detection of fake news in areas where they do not feel confident.

For this purpose, we conducted a study where participants identified fake news using an instrument of information quality (IQ). \citet{wang_beyond_1996} define IQ as information that is suitable for information users to make use of and \citet{gustavsson_assessing_2009} as the ability to satisfy the stated and implied needs of the information consumer. It is a multidimensional construct, consisting of several dimensions that describe certain aspects of IQ and can be measured and assessed \citep{gustavsson_assessing_2009, paggi_towards_2021}. We first build upon existing studies on assessing source documents using IQ as an instrument \citep{yaari_information_2011}. Second, we argue that fake news affects IQ and can be detected. Previous research has studied agreement between assessors when assessing IQ based on IQ dimensions and thus shown that IQ can be perceived and measured \citep{arazy_measurability_2011, fidler_improving_2017}. Following their findings, fake news might reflect on IQ dimensions, and it might be possible to detect them by the assessor.

Assessing IQ has proved to be difficult \citep{arazy_heuristic_2017}. Hence, fake news detection based on IQ might also be a difficult task. In this study, we sought to investigate how successfully users can detect fake news without the help of external tools and applications. Users assess the information that they are familiar with, with higher confidence than the information that they are unfamiliar with. Thus, in our study, we controlled for the influence of foreknowledge, because we argue that it is one of the vital determinants of fake news detection.

Existing studies have also investigated the positive correlation of formal education with observed variables (e.g.~the impact of formal education on lodging professionals' management success \citep{solnet_formal_2010} and the degree of payment inclusion \citep{polasik_impact_2018}). We assume that formal education might also impact fake news detection score. Thus we also examined the impact of formal education on the ability to detect fake news.

Besides users' core ability to detect fake news and familiarity with the source topics, personality traits \citep{gosling_very_2003} might also affect users' ability to detect fake news. Previous studies have investigated the relationship between personality traits and online content \citep{mehdizadeh_self-presentation_2010, amichai-hamburger_social_2010, ehrenberg_personality_2008, wilson_psychological_2010, correa_who_2010}. However, their findings may not apply to fake news detection abilities. The question arises as to whether and in what way personality traits affect the ability to detect fake news. Thus, we sought to expand the literature by investigating the relationship between personality traits and the ability to detect fake news. Our study is based on the widely accepted psychological Big Five personality trait model, which measures individuals' personality traits \citep{correa_who_2010} and provides a solid personality framework \citep{yin_reposting_2020}.

This paper contributes by examining whether end-users can identify fake news using IQ as an instrument. To the best of our knowledge, no other studies investigate the use of IQ instrument for fake news detection. Since little is known about fake news detection using IQ, we seek to focus the research community's attention in this area. By investigating the impact of fake news on individual IQ dimensions, this study provides steps for fake news detection and confirms that fake news can be identified on the basis of information distortion. Further, our study examines the impact of education, domain knowledge and personality traits on fake news detection score.

The remainder of the paper is organised as follows. In section \ref{related-work} we provide a review of related work regarding fake news detection, IQ instrument, and Big Five personality trait framework. In section \ref{problem-statement}, we introduce the problem statement and the research questions. In section \ref{experiment} the process of conducting the experiment and data collection is outlined, while in section \ref{findings}, empirical results, discussion, practical implications, and limitations are presented. Finally, in section \ref{conclusion}, we present conclusions, and suggestions for future work.

\hypertarget{related-work}{%
\section{Related work}\label{related-work}}

Fake news has spread all over the internet. It is often used to confuse people and persuade them to believe something, to buy a product or otherwise provide profit for the creator \citep{zhang_overview_2020}. People are usually not very good at detecting fake news \citep{pennycook_who_2020, moravec_fake_2018}, so it may become a powerful tool in the wrong hands \citep{allcott_social_2017}. This is especially true given that fake news spreads quickly, almost in real-time, data are far from verified, and it is generated in large volumes \citep{zhang_overview_2020}. Fake news stories spread faster and more widely than true news stories \citep{vosoughi_spread_2018}, and it is becoming a major concern that fake news is massively present on social media and the internet in general.

On social media, fake news contains four basic components: creator, target victims, news content, and social context \citep{zhang_overview_2020}. Creators want to influence the target audience through the content they publish on social media and persuade the audience to redistribute it. To stop fake news from spreading, creators should verify news before sharing it, and target users should verify content before sharing it. Since creators of fake news are often motivated for spreading it, the responsibility for detecting moves from the creator (publisher) to end-users.

Misinformation on social media has gained a lot of attention in recent years, which can be seen as an increase in the number of papers investigating fake news detection. Recent surveys have provided an extended review of approaches, datasets, automatic detection techniques, and potential future research directions \citep{bondielli_survey_2019, meel_fake_2020, zhou_survey_2020}. Recent studies \citep{lazer_science_2018} also highlighted the importance of multidisciplinary fake news research, an approach that we try to contribute.

In the field of fake news detection, much has been done in recent years. Notably, most of the false news detection approaches deal with the classification problem, where a particular piece of text has to be assigned in one of two sets (false, true) \citep{bondielli_survey_2019}. \citet{alrubaian_credibility_2018} presented a novel classification system that uses naive Bayes, random forest, J48, and feature rank models to distinguish between credible and non-credible content on Twitter. \citet{qin_predicting_2018} highlighted an innate shortcoming of rumour detection systems -- the ability to recognize rumours only when they have started spreading and causing harm. They introduced a novel rumour prediction approach that uses a Support Vector Machine (SVM) trained model.

Network structures are also used for automatic credibility assessment of target sources. \citet{vosoughi_rumor_2017} proposed an approach that uses network propagation dynamics and Hidden Markov Model (HMM) for preventing real-world false information on Twitter from going viral. \citet{ishida_fake_2018} used a bottom-up approach with a relative, mutual, and dynamic credibility evaluation using a dynamic relational network.

For false news detection, SVMs outperformed several supervised machine learning approaches according to findings in \citep{zhang_improving_2012}, however, the performance can variate depending on the training dataset. \citet{ruchansky_csi_2017} used a long short-term memory (LSTM) network as part of a framework for the task of fake news detection, while \citet{volkova_separating_2017} has evaluated both a recurrent neural network (RNN) and a convolutional neural network (CNN) approach. Many recent works have exploited a mixture of RNNs and CNNs in their models \citep{ajao_fake_2018, wang_liar_2017}. A recent approach \citep{sahoo_multiple_2021} employed temporal linguistic features and modified RNN to detect fake news. \citet{malhotra_classification_2020} used graphical convolutional networks and transformer-based encodings for fake news detection on Twitter. In their approach, \citet{ozbay_fake_2020} combined text mining techniques and supervised artificial intelligence algorithms.

In another line of research, \citet{pennycook_who_2020}, \citet{moravec_fake_2018} and \citet{zhu_what_2020} are more interested in end-users and their ability to detect fake news. The traits of regular social media users are worth exploring because social media has moved quality control for detecting fake news from trained journalists to regular users \citep{kim_says_2018}. \citet{jang_meaningful_2019} studied individuals' online sharing behaviour and found that a video with a meaningful or morally motivated content will be spread more virally than a video with primarily hedonic content. \citet{moravec_fake_2018} examined whether social media users can detect fake news on social media and whether a fake news annotation would help with this determination. They found that social media users are ineffective at separating fake from real news and a fake news annotation did not affect users' judgement about truth. They indicated that users may be easy victims of their opinions and believe what they want to believe. \citet{pennycook_who_2020} studied the psychological profile of individuals who fall prey to fake news. They conducted studies with \(1,606\) participants via Amazon's Mechanical Turk. They found that fake news may be fuelled by people who too easily believe unproven claims. They also found a positive correlation between individuals who overclaim their level of knowledge and more accurately judging fake news.

Online user reviews are an important form of electronic word-of-mouth communication \citep{hajli_social_2015}. The issue of fake reviews online is gaining relevance due to a growing number of cases of deceptive corporate practices and the importance of such reviews for consumers \citep{plotkina_illusions_2018}. \citet{zhu_what_2020} analysed how users perceive the quality of reviews. In the real world, binary classification of user reviews into real and fake may be too strict, so they employed perceived IQ as means for quality detection. \citet{zhu_how_2020} examined the relationship between online reviews and purchase intention by applying the stimulus--organism--response framework. In their approach, \citet{nakayama_exploratory_2017} used IQ to investigate whether fake (helpfulness) votes influence users' judgements of review IQ. However, they employ IQ as a unidimensional rather than a multidimensional construct. All these approaches highlight the importance of IQ content as a tool for fake news assessment online.

The sheer volume of fake news suggests automated online fake news detection approaches are desirable. However, such approaches face multiple challenges; among them, that the detection process is time-consuming, the results are always delayed, fact-checking tasks are still mainly dependent on human knowledge, and the limited availability of high-quality labelled datasets \citep{zhang_overview_2020}. An alternative approach is to focus more on a receiver (e.g.~reader) and its ability to detect information richness of a message. In our approach we assume that the IQ of fake news is negligible.

The literature on fake news is growing, but little has been done to incorporate IQ in fake news research. We argue that it is necessary to include an IQ perspective to understand fake news. The aforementioned researches by \citet{zhu_what_2020} and \citet{zhu_how_2020} used IQ to study users' perception of online reviews. Their findings indicate there is a link between information richness and perceived IQ.

The main question is whether IQ can be used for fake news detection. To the best of our knowledge, no existing research has studied this topic. Our research highlights the importance of a receiver (end-user) being able to detect fake news in content using IQ as a tool. We have investigated which IQ dimensions might allow fake news to be detected, and to which IQ dimensions users should devote more attention when assessing news from unverified sources.

Past research has studied cues that can be associated with lower receptivity to fake news. \citet{yin_reposting_2020} investigated how personality traits combined with negative emotions affect users' reposting of negative information on microblogs. \citet{pennycook_who_2020} investigated the psychological profile of individuals who accept fake news as true. \citet{mehdizadeh_self-presentation_2010} studied how narcissism and self-esteem are manifested on Facebook. \citet{amichai-hamburger_social_2010} highlighted personality traits that are important to social media: extraversion, neuroticism, and openness to experience. \citet{ehrenberg_personality_2008} found that individuals with high neuroticism and extravertion use instant messaging services more than others. \citet{wilson_psychological_2010} findings indicate that extroverted and unconscientious individuals reported higher levels of both social network site use and addictive tendencies. \citet{correa_who_2010} confirmed that extraversion and openness to experiences were positively related to social media use. Based on the results in past research, we can argue that personality traits may be inextricably linked to the perception of fake news and actions performed. We speculated that personality traits may have a significant impact on information evaluation, and were interested in how personality traits affect success in fake news detection using IQ as a tool.

\hypertarget{problem-statement}{%
\section{Problem statement}\label{problem-statement}}

As the volume of content grows, it is more and more difficult to acquire complete and accurate information. We argue that the approach of recognizing fake news based on IQ is worth exploring. To identify fake news based on the selected set of dimensions, we analysed short news of differing IQ quality and studied how individual IQ dimensions correlated with fake news. In this work, we highlight the relationship between IQ and susceptibility to fake news.

Only a few researchers have employed IQ in studies of fake news. \citet{agarwal_information_2010} investigated the relationship between \(16\) IQ dimensions and different types of social media. However, that study did not focus on how fake news affects IQ dimensions. \citet{zhu_what_2020} analysed the differences in users' perceived quality of online reviews on \(10\) diverse IQ dimensions, while a study from \citet{zhu_how_2020} examined the effect of perceived IQ on purchase intention. Neither of these studies focused on fake news. The aforementioned studies also used an extended range of IQ dimensions. Some previous studies that examined the impact of content on IQ dimensions \citep{arazy_measurability_2011, arazy_heuristic_2017} used a reduced set of dimensions (i.e.~accuracy, objectivity, representation and completeness) rather than attempting to cover the extended range. Like \citet{arazy_heuristic_2017}, we believe that this subset of dimensions reasonably represents different aspects of IQ. Furthermore, our use of a reduced set makes the study more directly comparable to previous work in the field.

People access online information from various sources that vary in terms of functionality and content length. We believe that content's length has an important impact on IQ and should not be neglected. In terms of functionality, \citet{agarwal_information_2010} proposed the following categorization of social media sites, which are arranged in a descending length order: wikis, social news, blogs, social networks, microblogging (tumblelogs), media sharing, and bookmarking. Wikis are publicly edited encyclopaedias and are usually longer texts. In general, blogs can be long, but usually contain fewer than \(2,000\) words in an individual blog post. Text posts on social networks (e.g.~Facebook) are much shorter than that. The optimal post length on social media is between \(40\) and \(60\) characters, while microblogs (e.g.~tweets) are further limited to \(160\) characters. \citet{lahuerta-otero_looking_2016} investigated the characteristics of influencers tweets and found that tweet length does not have a significant effect on microblogging users' influence. Wikipedia has been used several times as a source in previous IQ research \citep{luyt_improving_2010, fallis_toward_2008, lim_how_2009, arazy_measurability_2011, arazy_heuristic_2017}. A limited number of papers have investigated other categories from an IQ perspective. \citet{zhu_what_2020} and \citet{zhu_how_2020} studied online reviews. A typical online review length is from \(100\) to \(1,000\) words \citep{iio_evaluating_2012}, so we categorize these as blogs. In their studies, \citet{mai_quality_2013} and \citet{fidler_improving_2017} focused on documents that are shorter than wikis and typical blogs. They used a cooperative principle \citep{mai_quality_2013} to shorten original Wikipedia articles to make them more manageable for typical Wikipedia users (e.g.~students) while still retaining source IQ. In this study, we focused on an even smaller unit of text -- individual paragraphs. The paragraphs under study consist of a few sentences that refer to various IQ dimensions. We sought to investigate how users can identify fake news when dealing with paragraph-size source texts. The paragraphs are longer than posts on social networks and shorter than the documents studied by \citet{fidler_improving_2017}.

Several automatic approaches prevent false news from reaching end-users \citep{wu_misinformation_2019, zhou_survey_2020, viviani_credibility_2017}. While these approaches have several advantages, they do not necessarily isolate users from fake news. We believe that end-user detection capabilities should also be addressed and investigated. \citet{pennycook_who_2020} supported this idea -- claiming that it is important to understand the cognitive factors that allow readers to weed out the untrue in favour of the true. \citet{kim_says_2018} also stated that media has moved quality control for detecting fake news from trained journalists to regular users. Based on the idea that research should highlight the side of the information consumer, we investigated how successfully users can detect fake news based on the IQ carried by source documents.

Fake news detection using IQ is closely related to the end-user's (i.e.~news consumer) perception of the content. Previous studies have investigated how personality traits influence the perception of content. For example, \citet{pennycook_who_2020} investigated the psychological profile of individuals who fall prey to fake news. They suggested that there is a connection between the personal traits of users and receptiveness of users for pseudo-profound constructed fake news for garnering attention. The authors do not highlight the problem from an IQ point of view. Thus, our research aims to explore the impact of personality traits on the detection of fake news based on IQ dimensions.

In this study, we focused on determining whether it is possible to infer that some information is false based on information distortion. The same concept within the topic \(t\) was in our experiment described from the aspect of selected IQ dimension \(S\). A total of \(24\) statements per given topic \(t\) for each of the \(4\) IQ dimensions (accuracy, objectivity, representation and completeness) were constructed, resulting in half of the statements with a negative level of correctness \(l^{-}\) and the other half with a positive level of correctness \(l^{+}\) as further discussed in section \ref{experiment}. We conjectured that fake news is likely to reflect individual dimensions of IQ. We tried to detect fake news based on the quality of the information the media contained by user evaluation \(IQ_{init}(t,S,i)\) of a given statement from topic \(t\), from an IQ dimension \(S\) and in \(i\)-th story. To summarise, the input to our experiment was a randomly generated story, constructed of 4 short statements, each related to an IQ dimension \(S\) with a varying level of correctness \(l\). The output was the user's perception of IQ for a given story \(i\), i.e.~\(IQ_{init}(t,S,i)\). We stress that, while the literature on IQ focuses mainly on IQ value and its measurability, in this work, we investigated the lack of IQ as an indicator of misinformation.

Past research has shown that some IQ dimensions are easier to assess than others \citep{arazy_measurability_2011, arazy_heuristic_2017}. We sought to investigate if that also holds for fake news identification based on IQ dimensions. Furthermore, we also investigated which cues (e.g.~foreknowledge, education) affect fake news detection. We aimed to further explore the correlation between evaluators' personality traits and the perception of news as fake or genuine. In general, we studied whether IQ can successfully be used as a tool for fake-news detection.

Fake news detection using IQ is closely related to the end-user's (i.e.~news consumer) perception of the content. Previous studies have investigated how personality traits influence the perception of content. \citet{pennycook_who_2020} suggested that there is a connection between the personal traits of users and receptiveness of users for pseudo-profound constructed fake news for garnering attention. The authors do not highlight the problem from an IQ point of view. In this study, we investigate how personality traits affect fake news detection using the Big Five framework. We examine how personality traits represented by the Big Five personality dimensions affect fake news detection in terms of individual IQ dimensions. We measure personality traits using a custom, brief instrument based on a ten-item personality inventory (TIPI) \citep{gosling_very_2003}.

To the best of our knowledge, no prior study has empirically investigated IQ as a tool for fake news detection. We designed an empirical study to examine the fake news detection ability of different users and formulated the following research questions (RQ):

\begin{itemize}
\tightlist
\item
  \textbf{RQ1}: Are users able to perceive the IQ of short news and identify fake news?
\item
  \textbf{RQ2}: Which IQ dimensions make users more likely to detect fake news?
\item
  \textbf{RQ3}: Does education or domain knowledge affect users' fake news detection?
\item
  \textbf{RQ4}: Which personal traits facilitate fake news detection and which IQ dimension do these traits affect the most?
\end{itemize}

\hypertarget{experiment}{%
\section{Experiment}\label{experiment}}

\begin{figure}

{\centering \includegraphics[width=1\linewidth]{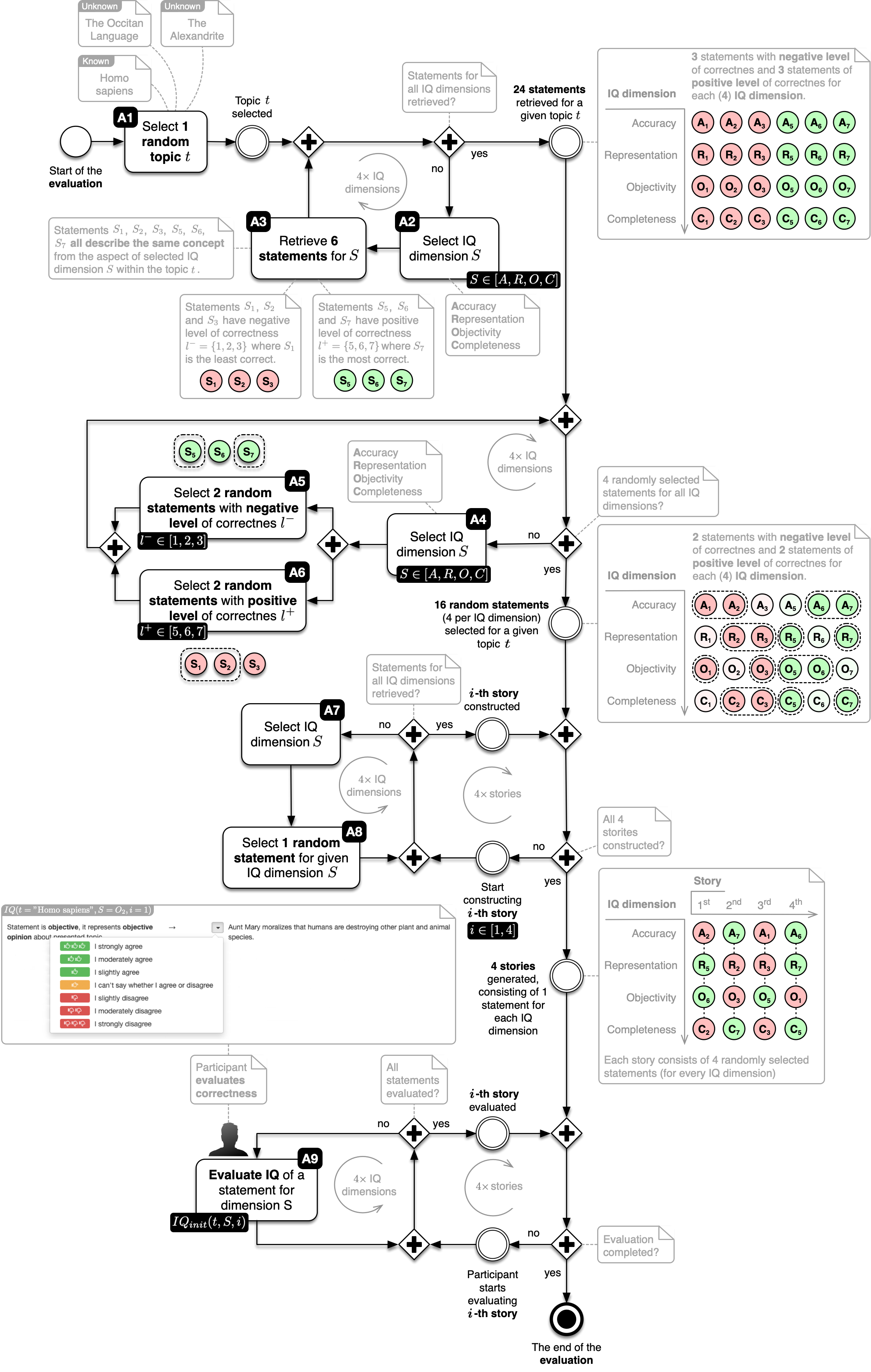} 

}

\caption{Steps in conducting an experiment}\label{fig:Approach}
\end{figure}

An online web application in a form of an interactive questionnaire was developed to explicitly support all steps of the experiment (see Figure \ref{fig:Approach}) and the link was distributed to \textbf{\(\boldsymbol{1123}\) participants} (\(\boldsymbol{31.6\%}\) female, \(\boldsymbol{68.4\%}\) male) from \(\boldsymbol{15}\) to \(\boldsymbol{69}\) years of age, with a median age of \(\boldsymbol{20}\). We targeted a population with various levels of education by directly inviting students from High School and University. To increase the diversity of participants and enhance the external validity of the research, we also reach out to other participants. The distribution of our sample was \(\boldsymbol{57.4\%}\) \textbf{undergraduate students} and \(\boldsymbol{42.6\%}\) \textbf{non-students}. The study focuses on the European region, where the participation in the study was voluntary with no paid compensation.

The questionnaire was divided into an introductory portion, with the collection of demographic information about participants, and the main portion that asked participants to evaluate several statements with varying level of correctness (inverse to a level of fakeness) on selected topics in terms of four IQ dimensions: objectivity, representation, accuracy, and completeness (Table \ref{tab:IQ-dimensions}). A similar study using a comparable approach to design the questionnaire and conduct the experiments was used in \citet{fidler_improving_2017}, \citet{lee_aimq_2002} and \citet{yaari_information_2011}.

\begin{table}

\caption{\label{tab:IQ-dimensions}Selected IQ dimensions adopted from related work and used in this study}
\centering
\begin{tabular}[t]{ll}
\toprule
Dimension & Description\\
\midrule
Accuracy & Information in the statement is accurate.\\
Completeness & Statement is complete and includes all necessary information.\\
Objectivity & Statement is objective; it represents objective opinion about presented topic.\\
Representation & Statement is presented consistently and formatted concisely.\\
\bottomrule
\end{tabular}
\end{table}

The motivation for conducting the research was presented to participants in the introductory portion of the questionnaire; they were then asked to provide some general information and their familiarity (knowledge level) with the topic of the statements being evaluated. Participants also carried out a self-assessment of their personality traits based on the Big Five personality model \citep{lamers_differential_2012}.

Several measuring instruments have been developed to measure the Big Five dimensions. In general, they are too lengthy for scenarios where participants' time is important \citep{woods_measuring_2005}. \citet{ehrhart_testing_2009} state that in such circumstances, an extremely short measure of the Big Five is adequate, and that researchers should consider the TIPI instrument \citep{gosling_very_2003}. We retained the same ten items from the TIPI instrument for measuring Big Five personality dimensions and asked participants to identify themselves in terms of positive and negative personality traits.

The average participant in our study was an individual with positive extraversion, neutral neuroticism, positive conscientiousness, positive agreeableness and positive openness to experience. The distribution of participants in Big5 groups was nearly uniform for Big5 factors neuroticism (\(35.3\%\) positive, \(35.8\%\) neutral and \(28.9\%\) negative), extraversion (\(41.3\%\) positive, \(31.7\%\) neutral and \(27\%\) negative) and conscientiousness (\(43.3\%\) positive, \(37.7\%\) neutral and \(19.1\%\) negative). Big5 factors agreeableness (\(57.3\%\) positive, \(33.2\%\) neutral and \(9.5\%\) negative) and openness to experience (\(49.1\%\) positive, \(41.9\%\) neutral and \(9\%\) negative) had a higher share of individuals within a positive Big5 group.

Algorithm 1 and Figure \ref{fig:Approach} illustrate all steps in the experiment process.

\begin{algorithm}[H]
\label{alg:1}
\DontPrintSemicolon
\SetAlgoLined
\BlankLine
\textbf{set} topics $\boldsymbol{T} = \{\ \text{the Occitan Language}, \text{Alexandrite}, \text{Homo sapiens}\ \}$ \\
\textbf{set} IQ dimensions $\boldsymbol{S} = \{\ A: \text{accuracy},\ R: \text{representation},\ O: \text{objectivity},\ C: \text{completeness}\ \}$ \\
\textbf{set} negative levels of correctnes $\boldsymbol{L^-} = [\ 1,\ 3\ ]$ \\
\textbf{set} positive levels of correctnes $\boldsymbol{L^+} = [\ 5,\ 7\ ]$ \\
\textbf{set} stories $\boldsymbol{I} = [\ 1,\ 4\ ]$
\BlankLine
\textbf{select} $1$ random topic $t \in T$
\BlankLine
\tcp{Retrieve 24 statements (6 for each S) for a given topic t}
\For{each IQ dimension S} {
  retrieve statements $S_1(t)$, $S_2(t)$ and $S_3(t)$ with $l \in L^-$ \\
  retrieve statements $S_5(t)$, $S_6(t)$ and $S_7(t)$ with $l \in L^+$
}
\BlankLine
\tcp{Select 16 random statements (4 for each S) for story construction}
\For{each IQ dimension S} {
  select $2$ random statements with $l \in L^-$ \\
  select $2$ random statements with $l \in L^+$
}
\BlankLine
\tcp{Construct 4 random stories consisting of 1 statement for each IQ dimension $S \in D$}
\For{each story $i \in I$} {
  \For{each IQ dimension S} {
    select 1 random statement
  }
}
\BlankLine
\tcp{Participant evaluates the correctnes of each story $i \in I$ and IQ dimension $S \in D$}
\For{each story $i \in I$} {
  \For{each IQ dimension S} {
    user evaluates information quality $IQ_{init}(t, S, i)$ of a statement $S$ for a given topic $t$ as part of a story $i$
  }
}
\caption{Steps in the experiment process}
\end{algorithm}

Every participant was randomly assigned one selected topic \(t\) from the pool of three available topics (one familiar and two unfamiliars to participants): \emph{\textbf{the Occitan language}}, \emph{\textbf{Alexandrite}}, and \emph{\textbf{Homo sapiens}} (see \textbf{step 6} in Algorithm 1 and activity \textbf{A1} in Figure \ref{fig:Approach}), as a source for the stories. Homo sapiens topic was one that participants were familiar with. Two other topics, with which the majority of participants were unfamiliar, were also selected. The lesser-known topic was either the mineral Alexandrite, an extremely rare gemstone, or the Occitan language, which fewer than \(1.5\) million people speak. Such a topic was intended to serve as a use case of how participants react when they come in contact with unknown, unverified information. Four different short stories were created from the selected topics for each participant, resulting in stories with the same topic, but IQ differing in each of the four dimensions.

For every selected topic \(t\) we prepared \(24\) statements with differing level of correctness (positive and negative) in terms of four IQ dimensions \(S\): objectivity, representation, accuracy and completeness (see \textbf{steps 7--10} in Algorithm 1 and Table \ref{tab:levelOfCorrectnes}). The source for our statement construction was Wikipedia as it is open for everyone and easily accessible. It is also used as a general source of information in the majority of information quality-related research \citep{fidler_improving_2017}. For each of the four IQ dimensions \(S\), statements \(S_i\) were retrieved from the body text of the selected topic \(t\) (see activity \textbf{A2} in Figure \ref{fig:Approach}). The dimensions are represented by \(S \in \{ A, R, O, C \}\), with \(A\) (accuracy), \(R\) (representation), \(O\) (objectivity), and \(C\) (completeness), while the statements of dimensions are denoted by \(S_i\). Then, the level of correctness for each statement \(S_i\) was adjusted. The procedure for adjusting the level of correctness is described in Table \ref{tab:levelOfCorrectnes}. It is important to emphasize that each sentence focuses only on the main aspect of the selected IQ dimension (accuracy with interval of values, completeness with the number of included details, objectivity with the credibility of actor conveying information and representation with format and style consistency of minimal and maximal numbers, including units). The dimensions are represented by \(S \in \{ A, R, O, C \}\), with \(A\) (accuracy), \(R\) (representation), \(O\) (objectivity), and \(C\) (completeness), while the statements of dimensions are denoted by \(S_i\). All selected statements chosen for individual dimension \(S\) describe the same concept. However, each of the statements \(S_i\) have a different level of correctness. More precisely, the first three statements \(\Big(\{ S_1, S_2, S_3 \}\Big)\) have negative levels of correctness \(l^{-}\), while the latter three \(\Big(\{ S_5, S_6, S_7 \}\Big)\) have positive levels of correctness \(l^{+}\) (see \textbf{steps 8--9} in Algorithm 1 and activity \textbf{A3} in Figure \ref{fig:Approach}). In Figure \ref{fig:Approach}, statements with a negative level of correctness \(l^{-}\) are colored red, while statements with a positive level of correctness \(l^{+}\) are colored green. The lower the index of the statement, the more negative level of correctness it is associated with -- that is, statement \(S_1\) is the least correct, while statement \(S_7\) is the most correct. The statement with the highest level of correctness \(l^{+} = 7\) was taken directly from the source. In the case of the topic \emph{Homo sapiens} and IQ dimension accuracy \(A\), this statement \(A_7\) was \emph{``The height of an average human measures from 150 cm to 180 cm''}. The following statement was then adjusted according to the rules presented in Table \ref{tab:levelOfCorrectnes}. Statement \(A_6\) \(\big(l^{+} = 6\big)\) decreased the range of accuracy by \(50\%\) and was defined as \emph{``The height of an average human measures from 157 cm to 173 cm''}. Statement \(A_5\) \(\big(l^{+} = 5\big)\) increased the upper limit to a sum of maximum value and \(50\%\) range and was defined as \emph{``The height of an average human is less than 195 cm''}. Statement \(A_3\) \(\big(l^{-} = 3\big)\) included only the upper limit and was defined as \emph{``The height of an average human is less than 180 cm''}. Statement \(A_2\) \((l^{-} = 2)\) decreased the upper limit to \(50\%\) of the maximum value and was defined as \emph{``The height of an average human is less than 90 cm''}. Statement with the lowest level of correctness \(A_1\) \(\big(l^{-} = 1\big)\) increased the upper limit to \(200\%\) of the maximum value and was defined as \emph{``The height of an average human is more than 360 cm''}. This procedure of statement construction was performed for every IQ dimension \(S\) and topic \(t\).

\begin{table}

\caption{\label{tab:levelOfCorrectnes}Details of \textbf{statement construction}. Statements with the level of correctness \(7\) are taken directly from the source, while statements with the level of correctness \(1\) are completely false.}
\centering
\begin{tabular}[t]{>{\centering\arraybackslash}p{1.5cm}>{\raggedright\arraybackslash}p{2.5cm}>{\raggedright\arraybackslash}p{2.5cm}>{\raggedright\arraybackslash}p{2.5cm}>{\raggedright\arraybackslash}p{2.5cm}}
\toprule
Level of correctness & Accuracy & Completeness & Objectivity & Representation\\
\midrule
7 & true range value [min, max] & true with all details & most credible actor & most consistent\\
\cmidrule{1-5}
6 & range is decreased by 50\% & least important detail removed & likely credible actor & added 'to' and space ' ' before measurement unit\\
\cmidrule{1-5}
5 & < max + 50\% range & most important detail removed & slightly credible actor & different number format\\
\cmidrule{1-5}
3 & < max & most important detail only & slightly unlikely credible actor & different number format and unformal measurement unit\\
\cmidrule{1-5}
2 & < 50\% max & 2 least important details only & unlikely credible actor & different number format and wrong measurement unit\\
\cmidrule{1-5}
1 & 2 x max & least important detail only & not credible actor & different number format, wrong numbers and measurement unit\\
\bottomrule
\end{tabular}
\end{table}

Once we created the statements with the full set of levels of correctness, we selected actual statements for the assessment process. In Figure \ref{fig:Approach}, this step is denoted by activities \textbf{A4}, \textbf{A5}, and \textbf{A6.} For each IQ dimension \(S\), we selected four random statements, two with a positive level of correctness (see \textbf{step 12} in Algorithm 1) and two with a negative level of correctness (see \textbf{step 13} in Algorithm 1). We selected statements from the set with positive \(l^{+}\) and set with negative statements \(l^{-}\) that were prepared earlier in the process. For example, in Figure \ref{fig:Approach} there are two randomly selected statements \(S_1\) and \(S_2\) with a negative level of correctness (see activity \textbf{A6}) and two random statements \(S_5\) and \(S_7\) with a positive level of correctness (see activity \textbf{A5}). The selection process was then performed for the remaining IQ dimensions \(S \in \{ A, R, O, C \}\) (see \textbf{step 11} in Algorithm 1). Up to this point in the process, we selected \(16\) random statements, four per IQ dimension, half with negative and half with positive levels of correctness, all on the same topic \(t\). These statements served as the source for creating the four short stories that participants assessed.

The last step before participants start the assessment process involved story construction. From the statements, prepared in the previous step, we created four short paragraph-size stories. In Figure \ref{fig:Approach}, this step is denoted by activities \textbf{A7} and \textbf{A8}. Each story consists of four randomly chosen statements, where each statement belongs to one IQ dimension (see example in Table \ref{tab:storyExample}). We created stories progressively. For the \(i\)-th story (see \textbf{step 15} in Algorithm 1), we first randomly selected an initial IQ dimension \(S\). If the \(i\)-th story did not include a statement for selected dimension \(S\) (see \textbf{step 16} in Algorithm 1), we selected a random statement from among that had not been included in some previous story (see \textbf{step 17} in Algorithm 1). Otherwise, we selected the next dimension from the set \(\{ A, R, O, C \}\). The story was complete when it contained exactly one statement from each IQ dimension -- a total of four statements. The level of correctness of particular statements in the constructed story was unknown to the assessor. When all the prepared statements were included in the stories, they were ready for the assessment.

\begin{table}[!h]

\caption{\label{tab:storyExample}An example of a generated story (see first story in Figure \ref{fig:Approach}).}
\centering
\begin{tabular}[t]{cl>{\centering\arraybackslash}p{1.5cm}>{\raggedright\arraybackslash}p{5cm}>{\raggedright\arraybackslash}p{4.5cm}}
\toprule
ID & Dimension & Level of correctness & Content of statement & Statement construction details\\
\midrule
A2 & Accuracy & 2 & The height of an average human is less than 90 cm. & Less than 50\% of maximum value, where average human measures from 150 cm to 180 cm.\\
\cmidrule{1-5}
R5 & Representation & 5 & An average human weighs from 55 kg - 83 kilograms. & Different number format, where most consistent form is 55 kg - 83 kg.\\
\cmidrule{1-5}
O6 & Objectivity & 6 & Environmentalists are warning that expansion of human population is affecting the environment and survival of other plant and animal species. & Environmentalists are likely credible actor in comparison to most credible actor - scientists.\\
\cmidrule{1-5}
C2 & Completeness & 2 & In general, the human body consists of arms and legs. & Two least important details only, where correct statement includes head, torso, pair of arms and legs.\\
\bottomrule
\end{tabular}
\end{table}

The final step of the experiment was the evaluation phase (see activity \textbf{A9} in Figure \ref{fig:Approach}), in which participants assessed the stories. Specifically, each participant rated four short stories (see \textbf{step 20} in Algorithm 1), each one paragraph long. Each paragraph is devoted to one topic and covers all four quality dimensions. Participants evaluated the correctness of statements in each paragraph (related to various IQ levels, see \textbf{step 21} in Algorithm 1) using a \(7\) point Likert scale (from strongly disagree (1) to strongly agree (7)). For every story, participants had to decide how strongly they agree with perceived IQ for selected dimensions as defined in Table \ref{tab:IQ-dimensions}.

\hypertarget{findings}{%
\section{Findings}\label{findings}}

\hypertarget{results}{%
\subsection{Results}\label{results}}

\begin{figure}

{\centering \includegraphics{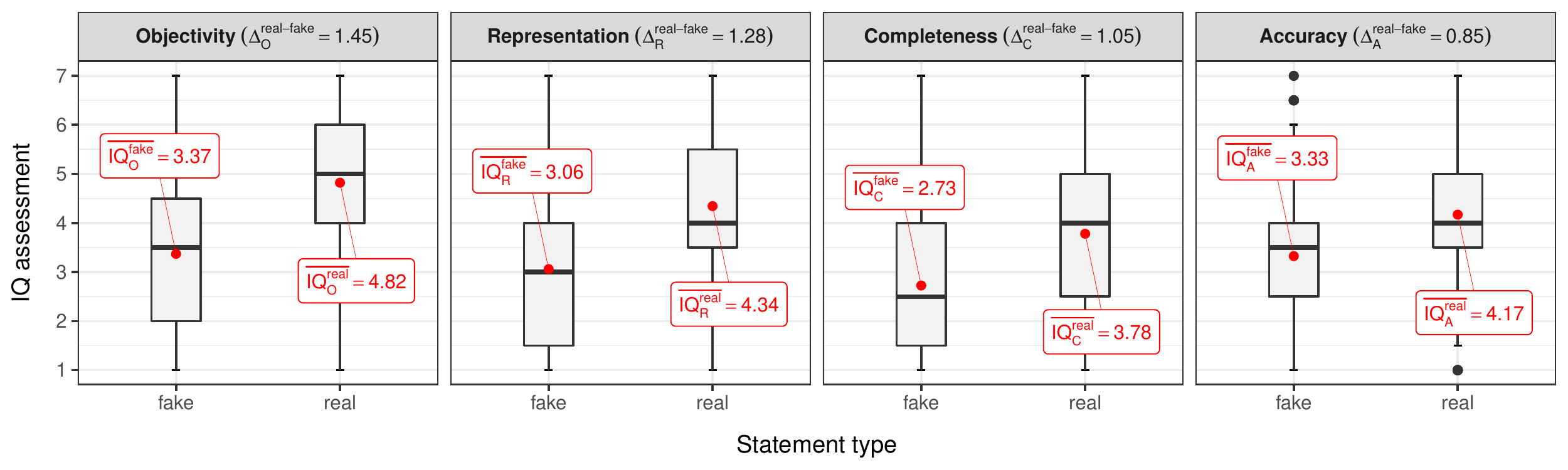} 

}

\caption{Successful \textbf{identification of fake and real statement types}}\label{fig:fake-real-by-IQ}
\end{figure}

As depicted in Figure \ref{fig:fake-real-by-IQ}, participants in our study successfully identified the differences between fake and real statement types throughout all considered IQ dimensions, although variations are significantly different across IQ dimensions.

To confirm that perceived \textbf{IQ assessment} \(\boldsymbol{IQ_d^t}\) is dependent on a \textbf{statement type} \(\boldsymbol{t} \in \{\text{fake}, \text{real}\}\) for every considered \textbf{IQ dimension} \(\boldsymbol{d} = \{ A, R, O, C \}\), we performed unpaired two-sample Wilcoxon or \textbf{Mann--Whitney test}s to compare two independent groups of samples. Our alternative hypothesis was \(\overline{IQ_d^{fake}} < \overline{IQ_d^{real}}\), with \(\alpha = 0.05\).

Participants were most successful in identifying the differences between fake and real information with the IQ dimension of \textbf{objectivity}, where the mean difference in perceived IQ assessments was calculated as \(\Delta_O^{\text{real} - \text{fake}} =\) \(\overline{IQ_{\text{O}}^{\text{real}}} - \overline{IQ_{\text{O}}^{\text{fake}}} =\) \(4.82 - 3.37 = \boldsymbol{1.45}\); the results are statistically significant \(\Big(W = \ensuremath{3.14\times 10^{5}}\), \(p = \ensuremath{3.52\times 10^{-95}}\Big)\). The effect of the remaining IQ dimensions (all with statistically significant differences) were as follows: \textbf{representation} with \(\Delta_R^{\text{real} - \text{fake}} = \boldsymbol{1.28}\) \(\Big(W = \ensuremath{3.5\times 10^{5}}\), \(p = \ensuremath{1.24\times 10^{-75}}\Big)\); \textbf{completeness}, \(\Delta_C^{\text{real} - \text{fake}} = \boldsymbol{1.05}\) \(\Big(W = \ensuremath{3.87\times 10^{5}}\), \(p = \ensuremath{1.51\times 10^{-57}}\Big)\); and \textbf{accuracy}, \(\Delta_A^{\text{real} - \text{fake}} = \boldsymbol{0.85}\) \(\Big(W = \ensuremath{4.18\times 10^{5}}\), \(p = \ensuremath{8.68\times 10^{-45}}\Big)\).

\begin{figure}

{\centering \includegraphics{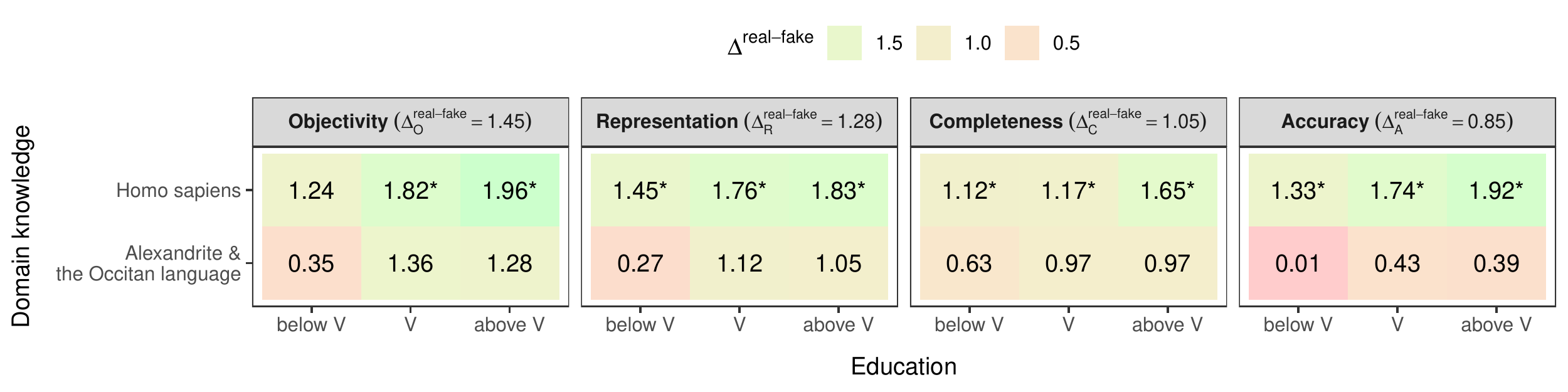} 

}

\caption{Influence of \textbf{education} and \textbf{domain knowledge} on identification of fake and real statement types. Domain knowledge includes two groups: Homo sapiens (well-known topic) and Alexandrite \& the Occitan language (less-known topic). Education is grouped into below V (levels 0--3, from early childhood, primary, to lower and and upper secondary education), V (level 4, post-secondary non-tertiary education) and above V (levels 5--8, from short-cycle tertiary, Bachelor's, Master's to Doctoral education), according to the ISCED 2011 classification.}\label{fig:education-by-domain-knowledge-graph}
\end{figure}

\begin{figure}

{\centering \includegraphics{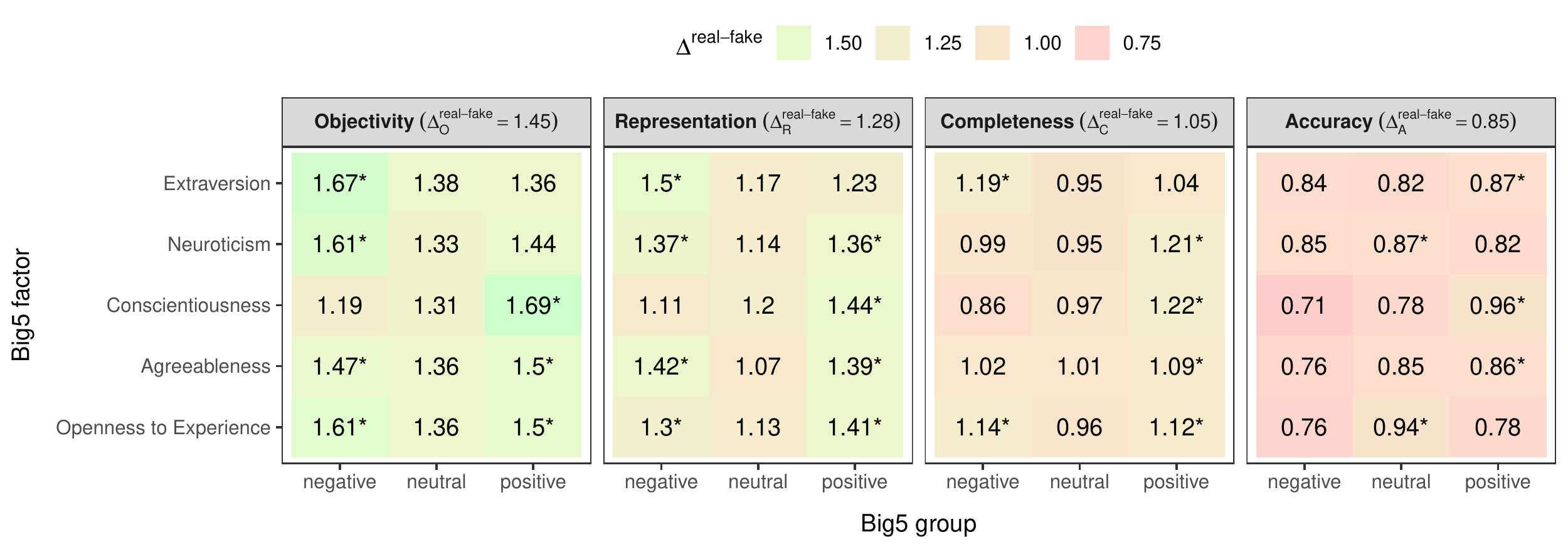} 

}

\caption{Influence of the \textbf{Big Five personality factors} on the identification of fake and real statement types}\label{fig:Big5-group-by-Big5-factor-graph}
\end{figure}

Figures \ref{fig:education-by-domain-knowledge-graph} and \ref{fig:Big5-group-by-Big5-factor-graph} depict the influence of the independent variables of education, domain knowledge, and Big Five personality factors. The values under investigation and depicted in the figures are \(\Delta^{\text{real} - \text{fake}}\). The impact of individual values regarding the mean value for a given IQ dimension is also marked with a coloured background.

Asterisks \textbf{(*)} in Figures \ref{fig:education-by-domain-knowledge-graph} and \ref{fig:Big5-group-by-Big5-factor-graph} depicts differences \(\Delta^{\text{real - fake}}_{d}\Big(\text{group}\Big)\) of \(\text{real}\) and \(\text{fake}\) statement types for a given IQ dimension \(d\), and \(\text{group}\), greater than the mean value \(\Delta^{real-fake}_{d}\) of the difference of \(\text{real}\) and \(\text{fake}\) statement types and a given IQ dimension \(d\).

\hypertarget{discussion}{%
\subsection{Discussion}\label{discussion}}

The purpose of this study was to empirically test the extent to which users (i.e.~information consumers) can identify fake news based on information distortion. We built our study on the assumption that fake news affects individual dimensions of IQ. To investigate this issue, we conducted a study in which participants assessed IQ in terms of selected dimensions of a source document in the form of short paragraphs. The dataset was carefully created according to the guidelines laid out by \citet{fidler_improving_2017}.

Regarding the first research question (\textbf{RQ1}), asking if users can perceive the IQ of short news and identify fake news, our results show that users are quite successful at identifying fake news. On average, they succeeded in detecting fake news on all IQ dimensions under study. The results depicted in Figure \ref{fig:fake-real-by-IQ} demonstrate that the difference between false and real statement types is greater than zero \(\big(\Delta^{\text{real - fake}}_d > 0\big)\) for all dimensions \(\big(\forall d \in \{A, R, O, C\}\big)\), where \(\Delta^{\text{real - fake}}_d = \overline{IQ^{real}_d} - \overline{IQ^{fake}_d}\).

The second research question (\textbf{RQ2}) focused on the detection capability of individual IQ dimensions. Our results show that there are major differences between IQ dimensions. While users are able to detect fake news more easily using some IQ dimensions, they may struggle with other dimensions. Participants were most successful in detecting differences in the IQ dimension objectivity \(\big(\Delta^{\text{real - fake}}_R = 1.45\big)\), followed by representation \(\big(\Delta^{\text{real - fake}}_R = 1.28\big)\), completeness \(\big(\Delta^{\text{real - fake}}_R = 1.05\big)\), and accuracy \(\big(\Delta^{\text{real - fake}}_R = 0.85\big)\).

Finally, the third (RQ3) and fourth (RQ4) research questions investigated the factors that influence fake news detection. We found that users' performance in fake news detection is influenced by several factors, such as knowledge of the problem domain, education, and users' personality traits.

\textbf{RQ3} deals with the impact of domain knowledge and education on the ability to more efficiently identify fake news. As depicted in Figure \ref{fig:education-by-domain-knowledge-graph}, users with knowledge of the problem domain \emph{(Homo sapiens)} were more successful in detecting fake news than users with deficient domain knowledge (Alexandrite and the Occitan language). A similar conclusion about the impact of knowing the problem domain was reached by \citet{xu_detecting_2019}. For the dimensions of representation, completeness, and accuracy \(\big(d_{RQ3} \in \{R, C, A\}\big)\), participants of our study achieved higher than average score \(\Delta_{d_{RQ3}}^{\text{real - fake}}\). For the dimension objectivity \(\big(d_{RQ3} = O\big)\), users achieved higher scores identifying fake news than average \(\Delta_O^{\text{real - fake}}\) if their level of education was V or greater. Only users with education lower than V struggled to successfully identify fake news, with scores \(\Delta_O^{\text{real - fake}}\) below average. In this case, foreknowledge itself is not enough, and more comprehensive knowledge is required to be successful in identifying fake news. We also observe that among users with foreknowledge of the problem domain, results for all four IQ dimensions \(d \in \{A, R, O, C\}\) consistently increase with a higher degree of education, meaning that education positively correlates with the ability to identify fake news. That is in line with the findings of \citet{deme_effect_2020}, who found similar effects concerning the impact of education on a different problem domain.

The fourth research question (\textbf{RQ4}) focused on which personal traits facilitate the detection of fake news and which IQ dimension is affected the most by these traits. As depicted in Figure \ref{fig:Big5-group-by-Big5-factor-graph}, we used the IQ instrument to measure the influence of Big Five personality traits on the ability to detect fake news across all IQ dimensions. For each personality trait, we examined how the positive and negative features of this human trait affect the ability to detect the IQ of content and thus fake news. If the result \(\Delta_{d_{RQ4}}^{\text{real - fake}} \big(f, g\big)\) for a given IQ dimension \(d_{RQ4} \in \{A, R, O, C\}\), Big Five personality factor \(f \in \{\) Extraversion, Neuroticism, Conscientiousness, Agreeableness, Openness to Experience \(\}\) and Big Five personality group \(g \in \{\) negative, neutral, positive \(\}\) was greater than the average value \(\Delta_{d_{RQ_4}}^{\text{real - fake}}\) for the IQ dimension \(d_{RQ4}\), then users with particular personal trait \(f\) and particular attitude of this trait \(g\) were more successful in detecting the difference among the IQ of real and fake statements. The background colour (from red to green) of \(\Delta_{d_{RQ4}}^{\text{real - fake}} \big(f, g\big)\) values in Figure \ref{fig:Big5-group-by-Big5-factor-graph} depicts the magnitude of a difference of the mean value \(\Delta_{d_{RQ4}}^{\text{real - fake}} \big(f, g\big) - \Delta_{d_{RQ_4}}^{\text{real - fake}}\).

The results in Figure \ref{fig:Big5-group-by-Big5-factor-graph} confirm that IQ dimension accuracy caused the most problems in identifying fake news, regardless of personality traits. Objectivity was the IQ dimension by which users most easily distinguished real from fake news, consistent with RQ2 and Figure \ref{fig:fake-real-by-IQ}.

As depicted in Figure \ref{fig:Big5-group-by-Big5-factor-graph}, individuals with pronounced \textbf{positive conscientiousness} \(\big(f^{\text{C}} =\) Conscientiousness and \(g^{+} =\) positive \(\big)\) were the most successful in identifying the differences between fake and real information \(\Delta^{\text{real - fake}}\), as they were able to identify fake news with all four IQ dimensions -- objectivity \(\big(\Delta_{\text{O}}^{\text{real - fake}} \big(f^{\text{C}}, g^{+}\big) = 1.69\big)\), representation \(\big(\Delta_{\text{R}}^{\text{real - fake}} \big(f^{\text{C}}, g^{+}\big) = 1.44\big)\), completeness \(\big(\Delta_{\text{C}}^{\text{real - fake}} \big(f^{\text{C}}, g^{+}\big) = 1.22\big)\) and accuracy \(\big(\Delta_{\text{A}}^{\text{real - fake}} \big(f^{\text{C}}, g^{+}\big) = 0.96\big)\). These results are higher than average values \(\Delta_{d_{RQ4}}^{\text{real - fake}}\) for each dimension \(d_{RQ4}\). Although users had some difficulties identifying fake news by IQ dimension accuracy, their result \(\Delta_{\text{A}}^{\text{real - fake}}\big(f^{\text{C}}, g^{+}\big)\) was still the highest among all groups \(\big(f, g\big)\) and above the average \(\Delta_{\text{A}}^{\text{real - fake}}\). According to the Big Five personality model \citep{gosling_very_2003, woods_measuring_2005}, conscientiousness is the personality trait of being honest, organized, and hardworking. \citet{sun_unique_2018}; \citet{maier_smartphone_2020} showed that among the five major factors, conscientiousness contributes the most to success in school and at work. Conscientious people usually achieve a higher level of education \citep{schneider_variables_2017}, which contributes to possessing more extensive knowledge; thus, they are more able to successfully detect misinformation. Our results confirm these findings from the literature that conscientiousness is positively correlated with education. The proportion of a group of participants with a pronounced positive conscientiousness (compared to neutral and negative groups) increased consistently from \(27\%\) for education group below V, to \(43\%\) for education group V and \(50\%\) for education group above V. This trend was also identified in domain knowledge, where the proportion of a group of participants with a pronounced positive conscientiousness increased from \(41\%\) for a low foreknowledge group (Alexandrite \& the Occitan language) to \(48\%\) for a high foreknowledge group (Homo sapiens). This influence of education was also identified in fake news detection in our RQ3. For all four IQ dimensions, there is also a growing trend in the ability to identify fake news among participants with a pronounced negative \(\big(\Delta_{d_{RQ4}}^{\text{real - fake}}\big(f, g^{-}\big)\big)\) conscientiousness compared to those with a pronounced positive \(\big(\Delta_{d_{RQ4}}^{\text{real - fake}}\big(f, g^{+}\big)\big)\) value (see Figure \ref{fig:Big5-group-by-Big5-factor-graph}). These findings confirm that conscientiousness facilitates the detection of fake news in terms of all four IQ dimensions.

The second most successful group of participants in fake news detection from results in Figure \ref{fig:Big5-group-by-Big5-factor-graph} were the \textbf{negative extroverts} or \textbf{introverts} \(\big(f^{\text{E}} =\) Extraversion and \(g^{-} =\) negative \(\big)\). They successfully detected the differences between fake and real information with IQ dimensions of objectivity \(\big(\Delta_{\text{O}}^{\text{real - fake}} \big(f^{\text{E}}, g^{-}\big) = 1.67\big)\), followed by representation \(\big(\Delta_{\text{R}}^{\text{real - fake}} \big(f^{\text{E}}, g^{-}\big) = 1.5\big)\), and completeness \(\big(\Delta_{\text{C}}^{\text{real - fake}} \big(f^{\text{E}}, g^{-}\big) = 1.19\big)\). Although \citet{ehrenberg_personality_2008} and \citet{correa_who_2010} found that extroverted people make more use of instant messaging services, our results show that the same group does not perform well at detecting fake news. Introverts prefer professions where they spend more time alone, e.g.~as artists, mathematicians, engineers and researchers \citep{sun_unique_2018, maier_smartphone_2020}. People who want to pursue such professions are usually enrolled in education for longer and gain more knowledge, much like conscientious individuals.

Individuals with pronounced \textbf{negative neuroticism} \(\big(f^{\text{N}} =\) Neuroticism and \(g^{-} =\) negative \(\big)\) were most successful at identifying fake news by the dimension objectivity \(\big(\Delta_{\text{O}}^{\text{real - fake}} \big(f^{\text{N}}, g^{-}\big) = 1.61\big)\). According to the Big Five personality model, these people are generally emotionally stable, confident, reliable, realistic, and practical \citep{sun_unique_2018, gosling_very_2003}. Interestingly, individuals with pronounced \textbf{positive neuroticism} \(\big(f^{\text{N}} =\) Neuroticism and \(g^{+} =\) positive \(\big)\) identified fake news quite successfully by the dimension completeness \(\big(\Delta_{\text{C}}^{\text{real - fake}} \big(f^{\text{N}}, g^{+}\big) = 1.21\big)\). According to the Big Five personality model, these are anxious people, complicating, emotional, sensitive, and not self-confident \citep{sun_unique_2018, gosling_very_2003}. \citet{ehrenberg_personality_2008} found that individuals with high neuroticism make more use of instant messaging services. Furthermore, authors in \citet{correa_who_2010} claim that in case of neuroticism, there is an association with instant messaging when compared to face-to-face interaction because instant messaging permitted additional time to contemplate responses. A similar conclusion was outlined by \citet{ryan_who_2011}, where individuals, who are high in neuroticism, are more likely to prefer using the Wall on Facebook, which offers people with neurotic tendencies the opportunity to take their time formulating messages and responses. Similarly, \citet{ma_impacts_2019} confirmed that neurotic people used LinkedIn for information purposes more, where they reacted more to professional information and also follow professional information more frequently. As individuals with pronounced positive neuroticism tend to use social media more, they process a higher share of content posted on social media than other users. Previous work has determined that they also take more time processing the content, which could explain their higher success rate in detecting false information by the dimension of completeness.

As depicted in Figure \ref{fig:Big5-group-by-Big5-factor-graph}, participants with both, positive and negative attitudes of Agreeableness and Openness to Experience achieved above-average results for the IQ dimensions of Objectivity and Representation. Thus, we can conclude that these two personality traits do not have a meaningful influence on the detection of fake news in terms of IQ dimensions of Objectivity and Representation. In the case of IQ dimensions of Completeness and Accuracy, participants with pronounced positive Agreeableness scored only slightly above average, but the trend from negative to positive attitude for both IQ dimensions is increasing, which leads to the conclusion that Agreeableness may also facilitate the detection of fake news via the IQ dimensions of Completeness and Accuracy.

Based on our results and regarding RQ4, we can conclude that participants with pronounced positive conscientiousness are the most successful in detecting fake news. They consistently outperformed in the detection of fake news using all four IQ dimensions. The second most successful were introverts, outperforming the average detection of fake news in three IQ dimensions, and the third most successful were negative and positive neurotics, in the IQ dimensions of objectivity and completeness, respectively.

Our findings may have important implications for today's world. Overall, research has found that the volume of fake news online is rising, and users are increasingly left to extract the truth themselves \citep{kim_says_2018}. In addition to intentional fabrication, sensationalism and competition for attention can also produce misinformation \citep{wang_systematic_2019}. However, our findings are not in line with those of \citet{moravec_fake_2018}. They found that social media users are poor at separating fake news from real news. Their work found that a fake news flag did not affect users' judgements about the truth when news challenged their opinions. We speculate that such a scenario holds for convinced users who are not interested in the truth, and we believe that such a scenario does not apply to our work.

Our findings can also be connected to recent research that individuals who overclaim their level of knowledge also judge fake news to be more accurate \citep{pennycook_who_2020}. The present study shows that education positively affects fake news detection. Previous work has shown that lack of knowledge is positively correlated with falling for fake news, and we showed that formal education improves the detection of false statements. The fact that domain knowledge has a huge impact on detection score is also in line with the findings of \citet{pennycook_who_2020}.

\hypertarget{practical-implications}{%
\subsection{Practical implications}\label{practical-implications}}

This study has several important implications for practice. First, our approach offers end-user an instrument for the manual assessment of shorter paragraphs for detection of misinformation. We believe that our approach can represent a valuable alternative to automatic approaches by exploiting different characteristics of false information. Automated approaches could have deficiencies such as a lack of high-quality labelled datasets, a lack of deep meaning, or inefficiencies in the early detection of misinformation \citep{shu_fake_2017, zhou_survey_2020}.

Second, information consumers must critically evaluate the information quality of online sources. Our study has shown that information consumer possesses the knowledge required to identify misinformation but need to be aware that critical appraisal is his/her responsibility. Our findings are also in line with those of \citet{wang_systematic_2019}, who stated that instead of retracting misinformation, equipping individuals with the faculty for a critical assessment of the credibility of information might yield better results.

Third, our results show that there are user groups among the general population that are more effective than average at detecting misinformation. We identified the personality traits that allow more effective manual detection of misinformation. That suggests we should consider the findings when selecting the right candidate for a job that requires manual detection of fake news. Based on this finding, we could greatly complement and improve existing systems for automatic fake news detection.

Fourth, we suggest greater caution when consuming news from different sources for users who do not have pronounced outlined personality traits or do not have a higher level of education, or are not familiar with the problem domain.

Fifth, in crowd-sourced fact-checking, assessment relies on a large population of fact-checkers that resemble end-users rather than expert-based checkers \citep{zhou_survey_2020}. We believe that our approach could empower crowd-sourced fact-checkers for better misinformation detection in a cost-effective manner.

\hypertarget{limitations}{%
\subsection{Limitations}\label{limitations}}

Although the current study contributes to previous research, it has some limitations. First, our study investigated paragraph-size documents, like those typically found in microblogs. We selected paragraph-size documents because microblogs are ideal platforms for information dissemination within a short period \citep{yin_reposting_2020}, and therefore, the diffusion of possible fake news. The effect of story size on IQ evaluation is addressed by \citet{fidler_improving_2017}, where they employed Gricean principles to shorten Wikipedia articles to paragraph-size documents. They compared agreement level and overall perceived IQ by the IQ dimensions used in our research. Their results confirmed that shortening the unit of evaluation from full-size text to paragraph-size did not influence the agreement level of evaluations.

We should also be aware of the limitations that come from participants who assess individual quality dimensions -- they may provide only a limited indication of the quality of the object \citep{arazy_measurability_2011}. This study focused on the detection of fake news by assessment of IQ. Such assessment is not an easy task, and \citet{arazy_heuristic_2017} showed that achieving agreement among assessors can be challenging.

We are also aware that we could use additional problem domains next to those employed in our study. However, creating statements with a specific amount of misinformation by individual IQ dimensions requires lots of effort. As part of our study, participants self-assessed their knowledge of the narrower and broader areas of the given problem domains. The results obtained were consistent with our assumption that the domain \emph{Homo sapiens} would be well known to them, while the other two were not.

Our study focused on fake news identification by considering various IQ dimensions. We employed a simplified method of acquiring information on participants' personality traits, requesting users perform a simple self-assessment. This approach is in agreement with the use of the TIPI instrument, but we are aware that more robust approaches exist (e.g.~BFI and NEO-PI-R). We did not utilize these comprehensive measurements of personality traits as they increase the payload to the participants who first have already agreed to invest substantial effort into a quality assessment of IQ dimensions.

\hypertarget{conclusion}{%
\section{Conclusion and future work}\label{conclusion}}

The responsibility for verifying content is increasingly being transferred from media sources to end-users on social media. We investigated how successful users are at detecting fake news in shorter content sources, namely paragraphs consisting of a few sentences. In terms of the factors that affect fake news detection, we examined the role of foreknowledge, education, and users' personality traits. Our approach used IQ as a tool for source document assessment, which allowed us to identify how participants perceive various IQ dimensions. Personality traits were evaluated using a custom instrument based on the established TIPI instrument. We constructed an online questionnaire and collected data from a variety of users.

The results demonstrate that we can use IQ-based tools to successfully detect fake news. We found that participants successfully identified the difference between fake and real statements according to all IQ dimensions studied. We also found that domain knowledge affects fake news detection. As far as education is concerned, we showed that higher education in combination with domain knowledge increases the ability to detect fake news. Without domain knowledge, education does not always contribute to performance in detecting fake news.

Concerning personality traits, we found that conscientiousness contributes significantly to better fake news detection across all IQ dimensions.

In future work, we would like to replicate our study on an extended body of participants. For instance, it would be beneficial to focus on participants considered established information experts, such as librarians, journalists, and professors. Tweets and other short sources are only one source of fake news on the internet. To address this issue, we would also like to extend our study to even larger source documents.

\hypertarget{acknowledgments}{%
\section*{Acknowledgments}\label{acknowledgments}}
\addcontentsline{toc}{section}{Acknowledgments}

We offer special thanks to Miloš Fidler, who was actively involved in the data collection process and contributed to the design in a manner that greatly assisted this research. We also thank all participants who generously shared their time for the purposes of data collection.

This research did not receive any specific grant from funding agencies in the public, commercial, or not-for-profit sectors.

\hypertarget{references}{%
\section*{References}\label{references}}
\addcontentsline{toc}{section}{References}

  \bibliography{references.bib}

\end{document}